\documentstyle[sprocl]{article}

\def\be{\begin{equation}}
\def\ee{\end{equation}}
\def\bea{\begin{eqnarray}}
\def\eea{\end{eqnarray}}

\begin{document}
\title{RECENT PROGRESSES OF ACCRETION DISK MODELS AROUND BLACK HOLES}
\author{ SANDIP K. CHAKRABARTI\footnote{e-mail:chakraba@bose.ernet.in} }
\address{S.N. Bose National Center for Basic Sciences, Salt Lake,
Calcutta 700091}
\maketitle\abstracts{
Accretion disk models have evolved from Bondi flows in the 1950s to
Keplerian disks in the 1970s and finally to advective transonic 
flows in the 1990s. We discuss recent progresses in this subject
and show that sub-Keplerian flows play a major role in determining the
spectral properties of black holes. Centrifugal pressure supported
enhanced density region outside the black hole horizon
produces hard X-rays and gamma rays by 
reprocessing intercepted soft photons emitted by the Keplerian disk 
terminated farther out from the black holes. 
Quasi-periodic oscillations can also be understood from the 
dynamic or thermal resonance effects of the enhanced density region.}
\section{Introduction}
Matter accreting on galactic and extragalactic black holes need
not be spherically symmetric Bondi flow or purely 
thin and Keplerian. In fact, since by definition,
matter must enter into a black hole with radial velocity on the
horizon similar to the velocity of light~\cite{{skc90},{skc96a}},
it must be supersonic and hence sub-Keplerian~\cite{skc96b}. If one
assumes that matter forms a thin, subsonic, Keplerian disk very far away
then it has to pass through at least one sonic point before the flow
enters the black hole. Second, since just outside of the 
black hole, the infall time scale is much shorter compared 
to the viscous time scale (even when viscosity is
high), angular momentum is roughly constant in the last few 
Schwarzschild radii and as a result, the centrifugal force increases
rapidly enough as the flow approaches the black hole so as to form a
centrifugal barrier behind which matter piles up. The resulting enhanced 
density region may be abrupt (if behind a shock), or smooth if the shock 
conditions are not satisfied. 
\section{Nature of Advective Solutions}
Chakrabarti~\cite{skc89} classified all possible solutions of the
inviscid adiabatic flows around black holes. Since angular
momentum is likely to be almost constant close to a hole even for
highly viscous flows, the inviscid solutions would be important even 
when viscosity is significant. It is observed that a large region of 
parameter space can produce standing shocks behind the centrifugal barrier. 
The classification is discussed in detail in Chakrabarti~\cite{skc96a}. 
In presence of viscosity, the advective disk may still produce 
shocks provided the viscosity parameter is less than some critical 
value~\cite{{skc90},{skc96b}}. The steady solutions with or without shocks 
in one and two dimensions have been tested to be stable by explicit numerical 
simulations~\cite{{mlc94},{rcm97}}. In a large region of the parameter 
space, two saddle type sonic points are present, but the shock conditions 
are not satisfied. The numerical simulations~\cite{rcm97} indicate that 
the shocks form nevertheless, but they are unstable, and 
oscillate back and forth. The period of oscillation depends on the specific 
angular momentum, but typically they can be around a fraction of a
second for galactic black hole candidates and around a day for extragalactic 
massive black holes. Even when stable shock conditions are  satisfied, shocks 
can be oscillatory when significant cooling effects are present~\cite{msc96} 
provided the cooling time scale in the post-shock region is comparable to 
the infall time scale in the pre-shock region. 

As viscosity is added, the closed topologies in the Mach No. vs. radial 
distance space of the inviscid solutions open up 
and the solution joins with a Keplerian disk farther out, provided
the accretion rate is large enough to have efficient emission from the
disk. For low enough viscosity and the accretion rate the
advective flow may join with a Keplerian disk very far away
(forming a giant primarily rotating ion torus which surround the disk of size
$10^{3-4} x_g$, see, g21-g41 topologies in Fig. 2a of 
Chakrabarti~\cite{skc96b}), while for higher viscosity
and accretion rate the Keplerian disk will come closer to the black hole
(see, g13-g14 topologies in Fig. 2a of Chakrabarti~\cite{skc96b}).
In intermediate viscosities, these flows may have shocks. For low
and intermediate viscosities the flow would definitely have a centrifugal
barrier and consequent enhanced density region in between the
black hole and the Keplerian disk. If viscosity varies with height,
it is expected that all the three types of flows would be manifested
in a single flow~\cite{{skclt95},{skc97}}. Chakrabarti~\cite{skc97}
discussed in detail the nature of the multi-component advective disk.
The boundaries of Keplerian and sub-Keplerian regions,
as well as the accretion rates in different components 
will vary from case to case as well as from time to time. As the viscosity 
at the outer edge increases, more and more matter goes from sub-Keplerian 
component to the Keplerian component. The soft photons from 
the Keplerian flow cools the sub-Keplerian component (thermal 
Comptonization) and hence the inner edge of the Keplerian 
component also advances. Eventually, the sub-Keplerian 
component cools completely and the Keplerian disk advances 
till the last stable orbit. Quasi-spherical flows in between 
the horizon and the last stable orbit reprocesses the 
soft photons out of the Keplerian component by transferring the
bulk momentum of the electrons to the photons~\cite{skclt95} 
(bulk motion Comptonization) and as a result, long extended power 
law component is formed even in the very soft state. 
\section{Spectral Properties}
When the accretion rate of the cooler Keplerian component (${\dot m}_d$)
is much smaller
compared to that of the hotter sub-Keplerian component (${\dot m}_h$), 
the soft photons emitted from the Keplerian component 
are unable to cool the hot electrons of the later component 
by thermal Comptonization processes. Thus, predominantly
hard X-rays are produced with little or no soft bump. Generally, this
soft bump may not be observed since the soft X-rays may be further
reprocessed by the extended atmosphere of the
disk. The X-ray spectral index $\alpha$ ($F(\nu) \sim \nu^{-\alpha}$) 
is around $0.5-0.8$ and ${\dot m}_d$ is typically less than 
$0.1-0.3$ when ${\dot m}_h=1$ depending on Models~\cite{skc97}.
As ${\dot m}_d$ is farther increased in comparison to  ${\dot m}_h$ 
(which may be due to sudden increase of viscosity in the flow 
[sudden capture of magnetic clouds from the companion, for instance] which
converts some of the sub-Keplerian flow into Keplerian~\cite{skcdm95}),
the soft photons of the Keplerian disk cools the electrons of the
sub-Keplerian component catastrophically and the spectra consists of only the
soft bump without any extended power law tail. This is called
soft state by some observers and this may happen for  Keplerian rate
of around $0.3-0.7$ or so when ${\dot m}_h=1$ depending on models.
One may also see broken power law in this state because
of the contribution from both thermal and bulk motion effects.
As ${\dot m}_d$ is farther increased, the
optical depth in the centrifugal pressure supported enhanced density region
becomes high and the flow drags most of the photons (which were supposed
to be emitted in between $3 x_g$ and $1 x_g$. The sub-Keplerian flow
farther out effectively looses its identity as it cools down 
completely to a temperature a little above the corresponding Keplerian
component~\cite{skclt95}). However, a fraction of
photons energized by infalling matter due to bulk motion Comptonization
can still come out. Chakrabarti \& Titarchuk~\cite{skclt95} (see
also Ebisawa et al.~\cite{etc96}
for details) showed that the power law hard component 
(with energy spectral slope $\alpha \sim 1.5-2.0$) can 
easily explain the behavior of the very soft state of the black hole
candidates. Recent computation~\cite{lgt97} shows that the power
law extends till almost $1$MeV. This power
law component is absent for both neutron stars as well 
as naked singularities~\cite{skcss97}. Thus, the presence of power law
component in very soft states has enabled observers to distinguish
a black hole from a neutron star very easily. Although the black hole
horizons permit energetic matter to be swallowed directly and therefore,
for a given accretion rate, black hole accretion could be less luminous
than the neutron star accretion~\cite{skcss97}, argument
based on total luminosity cannot be full proof, since there could
be any number of other physical effects (such as bipolar outflows 
which carry away energy and matter) confusing the situation.
\section{Quasi-Periodic Oscillations}
When Quasi-Periodic Oscillations (QPOs) were discovered in black hole
candidates a few years ago, it was surprising, since black holes have neither 
hard surfaces nor any anchored magnetic fields. It now appears, 
that the QPOs could be manifestation of the time-dependent solutions 
of the same set of equations which produced Keplerian and sub-Keplerian 
flows. As discussed in \S 2 above, simulations from a 
large region of parameter space showed that either because of 
resonance between inflow and outflow time-scales, or 
between inflow and cooling time scale, the 
enhanced density region oscillates with frequency very similar
to QPO frequency. Furthermore, these oscillating regions intercept
different amount of soft photons from the Keplerian component
and produces hard X-rays of significant amplitude as is observed.
Mechanisms based on acoustic oscillations or some such possibilities 
are incapable of producing significant modulations.

\section{Acknowledgments} The author thanks Goddard Space Flight
Center and USRA for partially supporting the cost of participation
at the conference.

\newpage

Send related preprints/reprints to the Author's Address:\\
\noindent  Prof. S.K. Chakrabarti
\noindent  S.N. Bose National Center for Basic Sciences
\noindent  JD-Block, Salt Lake, Calcutta -700091
\noindent  INDIA

\end{document}